# Self-organized Electronic Extended van Hove Singularity as Lattice Dynamic Confinement Effect

Sergei I. Mukhin

Theoretical Physics Department, Moscow Institute for Steel and Alloys, Moscow, Russia

A mechanism of self-organized one-dimensionality in correlated electron system coupled to optical phonon mode is proposed. It is found that a lattice vibration may compactify electron motion effectively to a one-dimensional space and trigger quantum phase transition into ordered state with "extended van Hove singularities" in the electronic Floquet modes spectrum. This mechanism may be of relevance for observed enhancement of the ordering instability in the anti-nodal regions of the "Fermi surface" in the high-$T_c$ cuprates, which is accompanied by anomalous softening of some optical phonon modes. A destruction of the effect by special microwave radiation is predicted, followed by a partial release of the zero-point vibration energy of the coupled optical phonon mode.

70.00.00, 71.38.+i, 71.35.Lk, 1115Kc

## INTRODUCTION

Though strong electron-phonon coupling in high-$T_c$ cuprates and parent compounds is now well documented experimentally[1,2], its role in the superconducting pairing leading to high superconducting transition temperatures in these materials is not well understood. Possibly, the central logical obstacle for this is that the Eliashberg theory of strong-coupling superconductivity may explain successfully the observed high values of $T_c$'s in cuprates only under condition that extended electronic van Hove singularities do exist.

These "anti-nodal" regions of high DOS pop-up *ad hoc* and must be directly involved in pairing[3]. The present paper is aimed to provide a missing link between the two ideas by proposing a novel mechanism of self-organized extended electronic "van Hove singularities" ("VHS") due to strong electron-phonon coupling in correlated electron systems. The quotes here are not misplaced and the purpose of the work is to explain this in a more detail.

The predicted phenomenon takes place in a form of a quantum phase transition into an ordered electronic state. This state and the lowest excited states of the electron-phonon system (separated at zero temperature with finite energy gaps) possess extended electronic "VHS" combined with finite amounts of the relevant optical mode softening and related discrete increments of its zero-point vibration amplitude. It is worth mentioning from the start that presented theory is neither about polaronic electron band-narrowing effect, nor about static crystal lattice modulation effect forcing electrons to move one-dimensionally. The theory describes self-organized redistribution of degrees of freedom in the many-body system, when a dynamic compactification of the electronic motion within one-dimensional space takes place at expense of the enhancement of zero-point vibration amplitude of the underlying lattice. The latter is accompanied by finite ($1/\sqrt{2}$ times) softening of the involved optical phonon mode. Even stronger, the actual one-dimensionality of the electronic DOS manifests itself not in the fermionic quasi-particle spectrum, but in the one-dimensional "nesting" of the Floquet indices of the electron states in the self-consistent time-periodic confining potential created by the relevant optical lattice vibration. The piling up of the Floquet indices of electrons coupled to a single optical mode (that also results in a strong softening of the latter) creates a

"saddle-point" in the relief of quantum action functional of the electron-phonon system. Appearance of the saddle-point signifies quantum phase transition into an extended electronic "Floquet-VHS" state mentioned above. There is also an interesting prediction that this state could be significantly modified by specially selected microwave electromagnetic radiation, that would simultaneously release some part of the zero-point optical phonon energy, while maintaining the one-dimensional electronic order (a unidirectional spin/charge density-wave). An influence of the external constant magnetic field on the considered confinement effect is also studied.

## PHASE TRANSITION THEORY WITHOUT ELECTRONIC QUASI-PARTICLE STATES

It is not surprising that theoretical description of the extended electronic Floquet-VHS necessitates some special formalism different from the common electronic band-theory approach. The derivation goes along the following path. The coupling of the electrons to an optical phonon mode being described by a Fröhlich Hamiltonian, for which a vacuum persistence amplitude $I$ is derived as the first step:

$$I = Tr e^{-iHt} = \int d\Psi d\bar{\Psi} dQ \times \exp\left[\frac{i}{\hbar}\int_0^t dt \int dV L\right] \equiv \int d\Psi d\bar{\Psi} dQ(t) \times \exp\left[\frac{i}{\hbar}S\right]. \quad (1)$$

Here $S$ is the action of the whole many-body system, described with the Hamiltonian:

$$H = H_e + H_{ph} + H_{eph},$$

that will be specified in detail below.

Now, using the well known equivalence between a many-body Hamiltonian and Lagrangian in the functional-integral representation, one has to integrate over hermitian-conjugated anti-periodic fermionic fields $\{\Psi(t), \bar{\Psi}(t)\} = \{-\Psi(0), -\bar{\Psi}(0)\}$ and periodic bosonic field $Q(t) = Q(0)$ representing the optical phonon mode. The (anti) periodicity of the (fermionic) bosonic fields in Eq. (1) ensures the equivalence between functional space integration and the trace operation[4].

Next, one calculates the (retarded) Green's propagator of the many-body system to find its eigen-energies spectrum:

$$G(E) = \sum_n \frac{1}{E_n - E} = iTr\int_0^\infty \frac{dt}{\hbar} \exp\left[\frac{iEt}{\hbar}\right] I(t) \equiv i\int_0^\infty \frac{dt}{\hbar} \int d\Psi d\bar{\Psi} dQ(t) \times \exp\left[\frac{i}{\hbar}(S + Et)\right] \quad (2)$$

Afterwards, the functional integration in Eq. (2) is made in the semiclassical approximation with respect to the phonon field $Q(t)$, that is determined self-consistently from the extremal $S$-action condition. Function $Q(t)$ enters as a time-periodic potential in the time-dependent Bogoliubov - de Gennes equations for the fermionic states. These fermionic states are characterized with the Floquet indices $\alpha_i$, that define phase-shifts of the electronic wave-functions acquired within one period of the time-periodic phonon field[4]:

$$\psi_i(\vec{r}, 0) = e^{-i\alpha_i} \psi_i(\vec{r}, t) \quad (3)$$

The Floquet index of the $i$-th quantum state in the time-periodic potential, defined in Eq. (3), is analogous to the quasi-momentum in a space-periodic static potential[4]. In the static, Hartree - Fock approximation, the Floquet indices trivially reduce to: $\alpha_i = t\varepsilon_i$, where $\varepsilon_i$ is the energy of the relevant quasi-particle stationary state. Notion of the fermionic Floquet indices spectrum permits in principle to integrate out fermions from Eq. (2) and derive exact expression for the Green's propagator in the form:

$$G(E) = \int_0^\infty \frac{dt}{\hbar} \int dQ(t) \times \prod_i \left( e^{i\frac{\alpha_i}{2}} + e^{-i\frac{\alpha_i}{2}} \right) \exp\left[\frac{iEt}{\hbar}\right] \equiv \int_0^\infty \frac{dt}{\hbar} \int dQ(t) \exp\left[\frac{i(S_{eff} + Et)}{\hbar}\right] \quad (4)$$

Here $S_{eff}(m, Q(t), t)$ is an effective action of the optical vibration mode. It has to be extremized with respect to the electronic order-parameter $m$ and lattice vibration $Q(t)$. Afterwards, the energies $E_n$ of the ground and excited states can be determined from the many-body analogue of the Bohr-Sommerfeld quantization condition[4]:

$$W(E) = S_{eff}(m, Q, t(E)) + Et(E) = 2\pi n\hbar; \quad n = 0, 1, \ldots, \quad (5)$$

where $t(E)$ is found from the stationary phase equation:

$$\frac{\partial S(t(E))}{\partial t} = -E. \quad (6)$$

This procedure also gives the renormalized frequencies and zero-point vibration amplitudes of the optical phonon mode in each state $E_n$.

As here we are interested only in the derivation of the energies of the ground and of the lowest excited states, the general expression in Eq. (4) can be farther reduced. Before coming to the main results it is worth to notice that since in general $\alpha_i \neq t\varepsilon_i$, we tackle here with a peculiar possibility of quantum suppression of the electronic qusi-particle states, because index $\alpha_i$ depends irreducibly on the (time-dependent !) optical vibration degree of freedom, and projected quasi-particle pole strength generally is not restricted to be non-vanishing.

## TRANSITION TO SELF-ORGANIZED ONE-DIMENSIONALITY

To outline the essentials of the novel effect the simplest possible model is considered of electrons in tight-binding approximation linearly coupled to a single lattice vibration

$\widehat{e}_\perp Q\cos(\Omega t)$, that exerts "deflecting" force $\widehat{e}_\perp gQ\cos(\Omega t)$ upon electrons along the axis $y$ perpendicular to the direction $x$ of the greatest electronic hopping amplitude $t_\parallel > t_\perp$. Here $g$ is the electron-lattice coupling constant and $\widehat{e}_\perp$ is a unit polarization vector. The Bogoliubov-de Gennes time-dependent Schrödinger equations take the form:

$$\left(i\partial_t \pm \mathbf{v}k + t_\perp(p) + i(gQ\cos(\Omega t) \pm \frac{e}{c}\mathbf{v}H + eE\cos(\omega_E t))\partial_p\right)\Psi_\pm - m^\pm \Psi_\mp = 0 \qquad (7)$$

where $\hbar = 1$, and the Lorentz force of the external (constant) magnetic field $H$ perpendicular to $x$-axis, as well as periodic external electric field force, $eE\cos(\omega_E t)$, are added for the future use. The rest of the notations are as follows: the $\pm$ sign reflects division of the states onto left- and right-movers, possessing quasi-momenta $p_x^\pm = k \pm k_F$ along the "longitudinal" hopping direction $x$, with $\mathbf{v}k \approx t_\parallel k/a$ being a linear part of the bare quasiparticle dispersion counted from the Fermi level, while $t_\perp(p) \approx t_\perp \cos p$ is the transverse part of the bare dispersion due to "free" quasi-particle hopping along $y$-axis. Hence, it is assumed that $t_\perp < t_\parallel$. The term with $\partial_p$ describes local lattice vibration acting on the electron in the transverse direction as "ping-pong paddles".

Periodicity of the phonon "orbits", imposed above and relevant for the trace operation, demands: $t = 2\pi/\Omega$. The (semiclassical) quantization of the lattice vibration is performed at the last stage by solving Eqs. (5) and (6). The bare frequency of the lattice vibration mode $\omega$ in combination with a proper atomic mass $M$ gives local lattice rigidity: $M\omega^2$. Solving Eq. (7) one finds Floquet indices according to Eq. (3) (in the case of zero external magnetic field and electric fields $H, E = 0$):

$$\alpha_{k,p} = J_0\left(\frac{gQ}{\Omega}\right)t_\perp \cos p \pm \sqrt{m^2 + (\mathbf{v}k)^2} \tag{8}$$

where $J_0(z)$ is the zero-th order Bessel function of the first kind. The $\{p,k\}$ components of the quasi-momentum represent the state index $i$ in Eq. (3). A remarkable consequence of Eq. (8) is that "ping-pong" transversal force exerted by the lattice causes collapse of the transversal dispersion $t_\perp(p)$ when the Bessel's function turns to zero:

$$J_0\left(\frac{gQ}{\Omega}\right) = 0, \quad \frac{gQ}{\Omega} = z_s, \quad z_1 \approx 2.405, \quad z_2 \approx 5.520, \quad \ldots \tag{9}$$

Here $z_s$ is a zero of the Bessel function $J_0(z)$. Physical meaning of this peculiar result is that a transversal periodic force acting on the electron dynamically confines its motion in the transverse directions and makes the motion effectively one-dimensional. This happens provided that the $p$-component of the electronic momentum traverses enough to overage out to zero the function $t_\perp(p)$ during one period of the lattice vibration. Substituting result from Eq. (8) into Eq. (4) and restricting contributions to the effective action as coming only from the ground and lowest excited energy states one arrives at:

$$S_{eff}(m,z,t) = \frac{t_\| t}{2}\left\{z^2\left(\left(\frac{2\pi}{\omega t}\right)^2 - 1 + \pi\left(\frac{m}{t_\|}\right)^2 \ln\frac{t_\|^2}{m^2 + t_\perp^2 J_0^2\left(\frac{t_\perp tz}{2\pi}\right)}\right)\right\}, \quad z \equiv \frac{Q}{a}, \tag{10}$$

where $a$ is the lattice constant, $M\omega^2 a^2 \approx t_\|$, and $g \approx t_\perp / a$, so that transverse electron hops are necessary for a scattering with $p$-component change. The "landscape" of $S_{eff}(m,z,t)$ as a function of dimensionless $\Omega/\omega$ at constant $m$ is presented in Fig. 1. It clearly demonstrates the presence of a saddle-point for a particular choice $z_s = z_1$. Extremizing Eq. (10) with respect to dimensionless vibration amplitude gives relation:

$z \approx z_m(t) = 2\pi z_s / gt \to 2\pi z_s / t_\perp t$. Substituting this relation into Eq. (10) and then solving Eqs. (6) and (5), one finds expressions for the spectrum of the electron-phonon system $E_n$, and related (renormalized) phonon frequency $\Omega_n$ and (dimensionless) zero-point vibration amplitude $z_n$ ($n = 0, 1, ...$).

## NONPERTURBATIVE OPTICAL MODE SOFTENING AND MAGNETIC FIELD EFFECT

Consider expressions for the energy of the ground state:

$$E_0 = \frac{t_\parallel}{2}\left(-\pi\left(\frac{m}{t_\parallel}\right)^2 \ln\frac{t_\parallel^2}{m^2} + \left(\frac{\omega}{t_\perp}\right)^2 \frac{z_s^2}{4}\right); \ \Omega_0 = \frac{\omega}{\sqrt{2}}; \ z_0 = \frac{\omega}{t_\perp}\frac{z_s}{\sqrt{2}} \propto z_\omega \left(\frac{\omega t_\parallel}{t_\perp^2}\right)^{1/2} \frac{z_s}{\sqrt{2}}. \quad (11)$$

The first term in the energy $E_0$ looks like a typical mean-field expression for the (quasi) one-dimensional electron system with $2k_F$ nesting vector between parallel flat pieces of the Fermi surface, that is characterized with the hopping integral $t_\parallel$ [5]. The enhancement of the zero-point lattice vibration energy in the second term in Eq. (11) reminds that the electronic one-dimensionality is *self-organized* property of the otherwise (anisotropic) 3D electronic system. The reason is that, in general, finite transverse hopping integral $t_\perp < t_\parallel$ would prevent the system from 1D-ordering. Indeed, adding inter-electron interaction energy in the mean-filed approximation to the first term in Eq. (11) and minimizing the sum with respect to order parameter $m$ one obtains minimal energy and non-zero equilibrium order parameter $m_0$:

$$E_m = \left(\frac{m^2}{t_\parallel U} - \frac{\pi m^2}{t_\parallel^2}\ln\frac{t_\parallel^2}{m^2}\right)_{min} = -\frac{\pi m_0^2}{t_\parallel} = \frac{\pi t_\parallel}{e}\exp\left\{-\frac{t_\parallel}{\pi U}\right\}. \quad (12)$$

The minimum exists provided that:

$$t_\perp \leq m_0 = t_\parallel \exp\{-t_\parallel/2\pi U\}/\sqrt{e}. \tag{13}$$

When bare $t_\perp$ violates quasi one-dimensionality condition in Eq. (13) the ordering transition with non-zero $m_0$ is, nevertheless, possible if phonon-induced rescaling of $t_\perp$ by a dynamic confinement effect is strong enough:

$$t_\perp \Rightarrow t_\perp J_0(gQ/\Omega) \leq t_\parallel \exp\{-t_\parallel/2\pi U\}/\sqrt{e}. \tag{14}$$

This means that interaction of electrons with the alternating phonon field effectively leads to the 1D-like "nesting" of the Floquet indices of electrons, expressed by Eqs. (8) and (14), making confining electron-phonon interaction the source of the electronic phase transition. Finite softening of the optical mode is expressed by the second relation in Eq. (11), it is manifestly nonperturbative. From Eq. (11) (see the last relation) it also follows that renormalization of the zero-point vibration amplitude $z_0$ as compared with the bare value $z_\omega$ is also nonperturbative. These findings correlate well with experimentally revealed strong coupling of the anti-nodal electronic states with momenta at $(\pi/a, 0)$ with $B_{1g}$ Cu-O bond-buckling long wave-length optical phonon[1,2] in high-$T_c$ cuprates. This effect may be related with known properties of the "anti-nodal" regions of the Fermi surface in these compounds. The regions look one-dimensionally flat ("holy cross"-feature) in the momentum distribution curves (MDC) obtained by energy-integrating of the ARPES spectral weight over a 500-mev window around the Fermi-level[2,7].

In magnetic field Eq. (10) is substituted with the following one:

$$S_{eff}(m,z,t) = \frac{t_{\parallel}t}{2}\left\{z^2\left(\left(\frac{2\pi}{\omega t}\right)^2 - 1\right) + \pi J_0^2\left(\frac{t_{\perp}}{\Omega_L}J_0\left(\frac{t_{\perp}tz}{2\pi}\right)\right)\left(\frac{m}{t_{\parallel}}\right)^2 \ln\frac{t_{\parallel}^2}{m^2}\right\}; \quad \Omega_L = \frac{eH}{m_e c} \quad (14)$$

where $\Omega_L$ is the Larmour frequency of electron. This leads to suppression of the lattice dynamic confinement effect introduced above in high enough fields $\Omega_L > t_{\perp}$, but in the lower fields $\Omega_L < t_{\perp}$ the lattice confinement effect dominates. From Eqs. (7) and (14) it follows immediately that effect of the external constant magnetic field actually may cause the same outcome, i.e. effective one-dimensionality of the electron motion. This magnetic field confinement effect was indeed proposed in the works by Gor'kov and Lebed' [5]. In the present case these two different mechanisms of the electron confinement compete with each other.

## MICROWAVE DETUNING OF THE LATTICE CONFINEMENT EFFECT

Even more pronounced and experimentally appealing is predicted below microwave-induced detuning effect of the self-organized electronic VHS. This detuning effect may in principle give a practical clue for a phonon-laser device. To see this we solve Eq. (7) with $H = 0$ and with non-zero alternating electric field $E$. Then, one finds modified Floquet indices using again Eq. (3):

$$\alpha_{k,p} = \left[\sum_{k=0}\left((-1)^k + (-1)^{kn}\right)J_k\left(\frac{gQ}{\Omega}\right)J_{kn}\left(\frac{eEa}{\omega_E}\right)\right]t_{\perp}\cos p \pm \sqrt{m^2 + (\mathbf{v}k)^2}; \quad n\omega_E = \Omega. \quad (15)$$

Here $a$ is a lattice constant, and commensurability relation between the frequencies $\omega_E$ and $\Omega$ is assumed: i.e. $n$ is positive integer. The factors in Eq. (15)

$J_{kn}(eEa/\omega_E) \propto \exp\{-kn/3\} \ll 1$ at $k > 0$, and the factor with $k = 0$ effectively switches off the electron-lattice dynamic confinement, provided that following condition is fulfilled:

$$t_\perp J_0\left(\frac{eEa}{\omega_E}\right) \leq t_\parallel \exp\{-t_\parallel/2\pi U\}/\sqrt{e}; \quad \frac{eEa}{\omega_E} \approx z_s \sim 1. \tag{16}$$

Under this condition (in units $\hbar = 1$) the Floquet indices spectrum in Eq. (15) again becomes *p*-independent, i.e. quasi one-dimensional. But, now this happens exclusively due to dynamic transverse confinement of electrons by microwave radiation, i.e. without invoking the self-consistently induced enhancement of the optical mode vibration amplitude. An estimate for the characteristic commensuration ratio: $n \gg 1$ follows from a request that the lowest necessary intensity *I* of microwave radiation obeying condition in Eq. (16) should be experimentally achievable:

$$I = \frac{c}{4\pi}E^2 \cdot 1\, cm^2 \propto \frac{c}{4\pi}\left(\frac{\hbar\Omega z_1}{nea}\right)^2 \sim 10^9\left(\frac{10^4}{n}\right)^2 \leq 10^8\, erg/s = 10W; \quad n \geq 10^4 \tag{17}$$

Here a characteristic optical phonon frequency $\Omega \sim 10^{13}\, \text{sec}^{-1}$ is assumed; $10W$ is taken for an experimentally achievable intensity of the microwave radiation, and the smallest zero root of the Bessel function $J_0$ is taken: $z_1 \approx 2.405$. At this large value of integer *n*, only $k = 0$ - term in the sum in Eq. (15) is significant. The latter fact makes condition in Eq. (16) sufficient for providing effective one-dimensionality of the Floquet indices spectrum. The effect of microwave localization of electrons is well known in molecular wires theory[1]. But in the present case the sufficient condition for microwave induced electronic ordering is the inequality in Eq. (16), that is much weaker condition than the equality $t_\perp J_0(eEa/\omega_E) = 0; (eEa/\omega_E) = z_s$[7] for a localization of electron in the wire with a monochromatic microwave radiation, which is difficult to achieve precisely using contemporary laser technique. Hence, condition for the microwave induced lattice deconfinement effect expressed in Eq. (16) is quite easy to check experimentally. Since, the increment of the lattice elastic energy of the zero-point vibrations, as given in Eq. (11), should be released at microwave detuning of the dynamic lattice confinement, this effect, besides fundamental interest, may be in principle important for creation of a phonon-laser.

## ACKNOWLEDGEMENTS

The author is grateful to Prof. A.A. Abrikosov for valuable discussion of results of this work prior to publication.

**FIGURE CAPTIONS**

Fig.1 The "landscape" of $S_{eff}(m,z,t)/\hbar$ as a function of dimensionless amplitude $z \in [0 \div 0.35]$ and frequency $\Omega \in [0.5 \div 0.9]$; $\Omega = 2\pi/t$. Other parameters entering $S_{eff}$, expressed in units of $t_\parallel$, are as follows: $m = 0.1$, $t_\perp = 0.4$, $\omega = 0.03$. The strongest saddle point is at $\Omega = 1/\sqrt{2}$; $z = 0.25$.

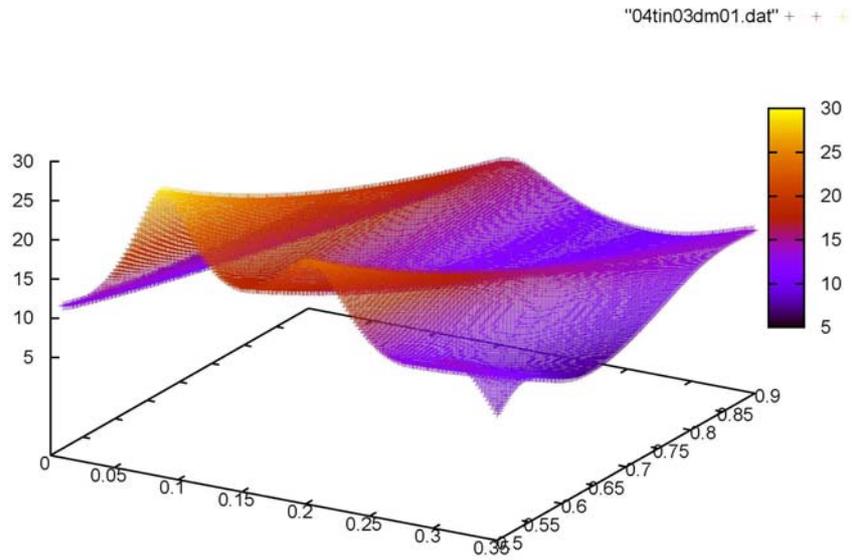

[1] J. Lee et al., Nature (London) **442**, 546 (2006).

[2] A. Lanzara et al., Nature (London) **412**, 510 (2001); T. P. Devereaux et al., cond-mat /, 0409426 (2004).

[3] A. A. Abrikosov, Physica C **244**, 243 (1995).

[4] R. G. Dashen, B. Hasslacher, and A. Neveu, Phys. Rev. D **12**, 2443 (1975).

[5] L. P. Gor'kov and A. G. Lebed', J. Phys. Lett. (Paris) **45**, L433 (1984).

[6] X. J. Zhou et al., Science **286**, 268 (1999).

[7] M. Grifoni, Phys. Rep. **304**, 229 (1998); J. Krczmarek, M. Stott, and M. Ivanov, Phys. Rev. A **60**, R4225 (1999).